\documentclass[final]{siamonline190516}
\usepackage{subfig}
\usepackage{amsmath}
\usepackage{balance}
\usepackage{lipsum}
\usepackage{amsfonts}
\usepackage{graphicx}
\usepackage{epstopdf}
\usepackage{algorithmic}
\DeclareMathOperator*{\argmin}{arg\,min}
\ifpdf
  \DeclareGraphicsExtensions{.eps,.pdf,.png,.jpg}
\else
  \DeclareGraphicsExtensions{.eps}
\fi
\usepackage{enumitem}
\setlist[enumerate]{leftmargin=.5in}
\setlist[itemize]{leftmargin=.5in}

\newsiamremark{remark}{Remark}
\newsiamremark{hypothesis}{Hypothesis}
\crefname{hypothesis}{Hypothesis}{Hypotheses}
\newsiamthm{claim}{Claim}

\headers{Street-level Travel-time Estimation via Aggregated Uber Data}{K.Maass, A.Sathanur, A.Khan, and R.Rallo}

\title{Street-level Travel-time Estimation via Aggregated Uber Data}
\author{Kelsey Maass\thanks{Pacific Northwest National Laboratory, Richland, WA 
  (\email{kelsey.maass@pnnl.gov})}
\and Arun V. Sathanur\thanks{Pacific Northwest National Laboratory, Seattle, WA 
  (\email{arun.sathanur@pnnl.gov})}
\and Arif Khan\thanks{Pacific Northwest National Laboratory, Richland, WA 
  (\email{ariful.khan@pnnl.gov})}
  \and Robert Rallo\thanks{Pacific Northwest National Laboratory, Richland, WA 
  (\email{robert.rallo@pnnl.gov})}}

\begin{document}
\maketitle

\begin{abstract}
Estimating temporal patterns in travel times along road segments in urban settings is of central importance to traffic engineers and city planners. In this work, we propose a methodology to leverage coarse-grained and aggregated travel time data to estimate the street-level travel times of a given metropolitan area. Our main focus is to estimate travel times along the arterial road segments where relevant data are often unavailable. The central idea of our approach is to leverage easy-to-obtain, aggregated data sets with broad spatial coverage, such as the data published by Uber Movement, as the fabric over which other expensive, fine-grained datasets, such as loop counter and probe data, can be overlaid. Our proposed methodology uses a graph representation of the road network and combines several techniques such as graph-based routing, trip sampling, graph sparsification, and least-squares optimization to estimate the street-level travel times. Using sampled trips and weighted shortest-path routing, we iteratively solve constrained least-squares problems to obtain the travel time estimates. We demonstrate our method on the Los Angeles metropolitan-area street network, where aggregated travel time data is available for trips between traffic analysis zones. Additionally, we present techniques to scale our approach via a novel graph pseudo-sparsification technique. 
\end{abstract}

% REQUIRED

% REQUIRED

\section{Introduction}
Data-driven mobility modeling and prediction are important aspects of modern urban planning. With respect to travel forecasting, the two major areas of research are demand modeling and travel-time estimation \cite{beimborn2006transportation}, where demand modeling involves generating accurate statistics of the number of trips between origin--destination (O--D) pairs, and travel-time estimation involves predicting the travel times for trips between O--D pairs. The focus of this work is on the latter, specifically, street-level travel-time estimation. In existing research on travel-time estimation, interstate link models have received disproportionate attention from the transportation research community, due primarily to the availability of large amounts of freeway sensor data. Although equally important, the same is not true for arterial models, where coverage is limited due to the costs related to installing probe sensors and associated infrastructure. Under these circumstances, significant insights can be gained at a fraction of the cost with coarse-grained data sets  such as the summary statistics of Uber trips.

Uber Movement datasets \cite{UberData} provide anonymized, aggregated, and coarse-grained O--D travel time statistics at the TAZ (traffic analysis zone) level for many metropolitan areas around the world. TAZs are small geographical units into which a given metropolitan area is divided, characterized by factors such as the total population, type of population, and employment. While Uber Movement datasets can be coarse-grained and have high uncertainty, they cover large geographical regions and are available for multiple metropolitan areas, allowing for generalizability. Similar datasets are also available through other sources (for specific cities) such as the New York City taxi-cab data \cite{NYCData}.

 Our work is concerned with filling the gap in arterial travel-time estimation. While our prior work \cite{Sathanur2019} focused on travel-time estimation at the TAZ level, this work is concerned with travel-time estimation at the street level. We start with the graph representation of a given road network, where intersections and road segments are represented as vertices and edges, respectively. Next, using the Uber Movement data and the street network graphs as inputs, we iteratively estimate the travel time on each edge for a given time window. At each iteration, we solve a constrained least-squares problem on the pseudo-sparsified graph. While the examples here utilize the coarse-grained, high-coverage Uber Movement data, the developed approach could seamlessly incorporate high-quality, low-uncertainty datasets, such as those from loop counters or probe sensors, by including these elements as constraints. While this work uses methods similar to those in \cite{bertsimas2019travel}, which combines shortest-path routing with a convex optimization formulation, we make the following contributions:
 
 \begin{itemize}
     \item We utilize trip sampling in order to leverage coarse-grained, aggregated TAZ-level data. Finer-grained data from sensors as well as street-level trip data can easily be absorbed into our approach.
     
     \item We propose a biasing scheme for sampling travel times from the statistics of the aggregated data, which can significantly improve predicted travel time distributions.
     
     \item We make use of constraints and the convex combinations of sequential iterates as a means of stabilizing solutions and improving convergence rates. 
     
     \item We demonstrate the efficacy of a graph pseudo-sparsification technique that can improve scalability with little loss of accuracy. We are currently building on this approach to scaling to much larger network sizes that encompass entire metropolitan areas with the aid of high performance computing resources. 
 \end{itemize}

The paper is organized as follows. Section \ref{sec3} describes the Uber Movement and the LA road network data used in this work. In the subsequent section (Section \ref{sec4}), we describe the forward model and simulated trips, followed by the optimization methodology (Section \ref{sec5}) used in this work. Section \ref{sec6} presents our principal experimental results, followed by the methods and scaling results related to our pseudo-sparsification technique in Section \ref{sec7}. Section \ref{sec2} describes related work in the area of graph analytics, optimization, and machine learning applied to road networks. We conclude the paper with a note on future work in Section \ref{sec8}.

\section{Datasets}
\label{sec3}
Our primary data sources for this work are from Uber Movement \cite{UberData} and the road network for the LA area. Uber has released a trove of aggregated and anonymized data on travel-time and average-speed statistics for a large number of cities around the world \cite{UberData}. Because gathering transportation data is an expensive and cumbersome process, leveraging these surrogate datasets is expected to help researchers and city planners conduct quick and fairly detailed studies of the various aspects of vehicle mobility in urban settings. In this work, we focus on Uber's travel time data, which includes statistics for travel times between pairs of TAZs or census tracts for each hour of the day and day of the week. 

The graph representation of the LA city road network is made available to us as part of the project. Similar analyses is possible with the open-source version of the maps provided by OpenStreetMaps \cite{OpenStreetMap}. Table \ref{tab1:comp_table} lists some basic network properties of the full LA road network, along with sub-networks formed with different radii around downtown LA. Other than the number of TAZs, most of the structural properties are similar across all of the networks. The number of TAZs included increases with the size of the graph.

\begin{table}[htbp]
\small
\resizebox{\textwidth}{!}{%
\begin{tabular}{|c|c|c|c|c|c|c|c|}
\hline
\textbf{Area} & \textbf{Vertices} & \textbf{Edges} & \textbf{TAZ} & \textbf{min(Deg)} & \textbf{max(Deg)} & \textbf{avg(Deg)} & \textbf{Clust. Coeff.} \\ \hline
\textbf{LA\_DT+1} & 3239 & 7138 & 25 & 2 & 12 & 2.2 &0.018 \\ \hline
\textbf{LA\_DT+2} & 9879 & 22756 & 79 & 2 & 12 & 2.3 & 0.025 \\ \hline
\textbf{LA\_DT+3} & 16906 & 40929 & 159 & 2 & 12 & 2.4 & 0.026 \\ \hline
\textbf{LA\_full} & 368419 & 905622 & 2205 & 2 & 14 & 2.46 & 0.034 \\ \hline
\end{tabular}%
}
\caption{\small Structural properties of LA road networks of varying sizes: one, two and three miles radii of downtown LA (DT) and the full LA network.}
\label{tab1:comp_table}
\vspace{-5mm}
\end{table}

\section{The graph-based forward model}
\label{sec4}
In this section we describe a forward model for travel-time prediction, and in the following section we show how the parameters of this model can be estimated by means of an optimizer. Let $\mathcal{P}$ denote a set of edges that forms a path from vertex $i$ to vertex $j$ in the road network graph. The expected travel time $y_{ij}$ between the vertices $i$ and $j$ can be computed as
\begin{equation}
y_{ij} = \sum_{k \in \mathcal{P}} t_k = s^T t,
\label{equn1}
\end{equation}
where $t_k$ represents the expected travel time along edge $k$, the vector $t \in \mathbb{R}^M_{>0}$ represents the travel time along all $M$ edges in the graph, and the binary vector $s \in \{0,1\}^M$ encodes the edges in $\mathcal{P}$. The choice of optimal routing ($\mathcal{P}$) between the origin and destination vertices is a variable in the model. Routing is typically done with special routing software such as the Open Source Routing Machine \cite{luxen-vetter-2011}. However, in this work, we use shortest-path routes where the edges are weighted by travel times \cite{bertsimas2019travel,lu2008mining,wu2012shortest}. Specifically, for initialization, we use weights determined by free-flow travel times $f \in \mathbb{R}^M_{>0}$, the time it takes to travel along a road segment free of congestion, computed by dividing the length of the road segment by the posted speed limit. 

\subsection{Training and testing tasks}
Eq.~\ref{equn1} is based on O--D pairs in the road network graph. However, the Uber Movement data is provided at the much coarser granularity of TAZ O--D pairs, consisting of travel-time statistics computed over all trips originating at a given TAZ and ending in another during a particular hour of the day. Furthermore, not all TAZ pairs are included in the data. 

For a given hour of the day and geographic area, we first collect all the TAZ O--D pair statistics available from Uber Movement. We then choose a random 90-10 split of the available TAZ O--D pairs for training and testing, respectively. For each of the TAZ O--D pairs, we simulate trips by sampling vertices from the origin and destination TAZs to form vertex O--D pairs, assigning trip times by sampling from a log-normal distribution based on the geometric mean and geometric standard deviation travel times given in the Uber dataset. 

After estimating edge travel times using our vertex-level training data, testing is done at the TAZ level using the geometric mean travel time for all vertex O--D pairs present in a given test TAZ O--D pair. For example, suppose there are $n_{ij}$ simulated trips from TAZ $i$ to TAZ $j$. The estimated geometric mean travel time would then be computed as
\begin{equation}
g_{ij}(t) = \left( \prod_{k = 1}^{n_{ij}} s_k^T t \right)^{\frac{1}{n_{ij}}},
\label{eqgmtt}
\end{equation}
where vector $s_k \in \{0,1\}^M$ encodes the edges in the weighted shortest-path between sampled vertex O--D pair $k$. The vertex O--D sampling and travel-time sampling are described in detail below.

\subsubsection{Vertex sampling}
\label{sec:vertsamp}
For each iteration $k$ of our edge travel-time estimation algorithm, we sample $N$ vertex O--D pairs each from our training and test sets, letting the number of simulated trips for each TAZ O--D pair be proportional to the size of the two TAZs. Specifically, for each TAZ O--D pair $(i,j)$ in the current subset $\mathcal{U}_k$ of the Uber dataset, we let $m_{ij}$ be the product of the number of vertices in origin TAZ $i$ and destination TAZ $j$, then we set the number of simulated trips $n_{ij} = \text{int}\left(\left(\frac{m_{ij}}{\sum_{k,\ell\in \mathcal{U}_k}m_{k\ell}}\right)N\right)$.

We sample the $n_{ij}$ origin and destination vertices for a given TAZ O--D pair uniformly from all vertices within their respective TAZ. We note that this process selects the shortest-path edges with a probability proportional to their local betweenness centrality, which in turn correlates with the importance of the edge with respect to traffic flow \cite{crucitti2006centrality,jiang2009street,porta2006network}. 

\subsubsection{Travel-time sampling}
Let $S \in \{0,1\}^{M \times N}$ represent the free-flow shortest-path matrix for our $N$ simulated trips. We assign travel times $y \in \mathbb{R}^N_{>0}$ to trips based on the ordering of free-flow shortest-path travel times $y_f = S^Tf$ for all vertex O--D pairs within a TAZ O--D pair, summarized in Algorithm~\ref{alg1} below. Here $f \in \mathbb{R}^M_{>0}$ denotes the vector of free-flow travel times for each of the $M$ edges. 
\begin{algorithm}
\small
\caption{Travel-time sampling}
\textbf{Input}: Free-flow shortest-paths matrix $S_k$ \\
\textbf{Output}: Sampled travel times vector $y_k$
\begin{algorithmic}
\FOR{all TAZ O--D pairs $(i,j) \in \mathcal{U}_k$}
    \STATE{Sample $n_{ij}$ travel times $y_s$ from log-normal distribution}
    \STATE{Get indices $\ell_s$ and $\ell_f$ of longest to shortest travel times for $y_s$ and $y_t$, respectively}
    \FOR{$m = 1,\dots,n_{ij}$}
        \STATE{Assign $y_k(\ell_f(m)) = y_s(\ell_s(m))$}
    \ENDFOR
\ENDFOR
\end{algorithmic}
\label{alg1}
\end{algorithm}

\begin{figure}[htbp]
  \centering
  \includegraphics[width=0.9\linewidth]{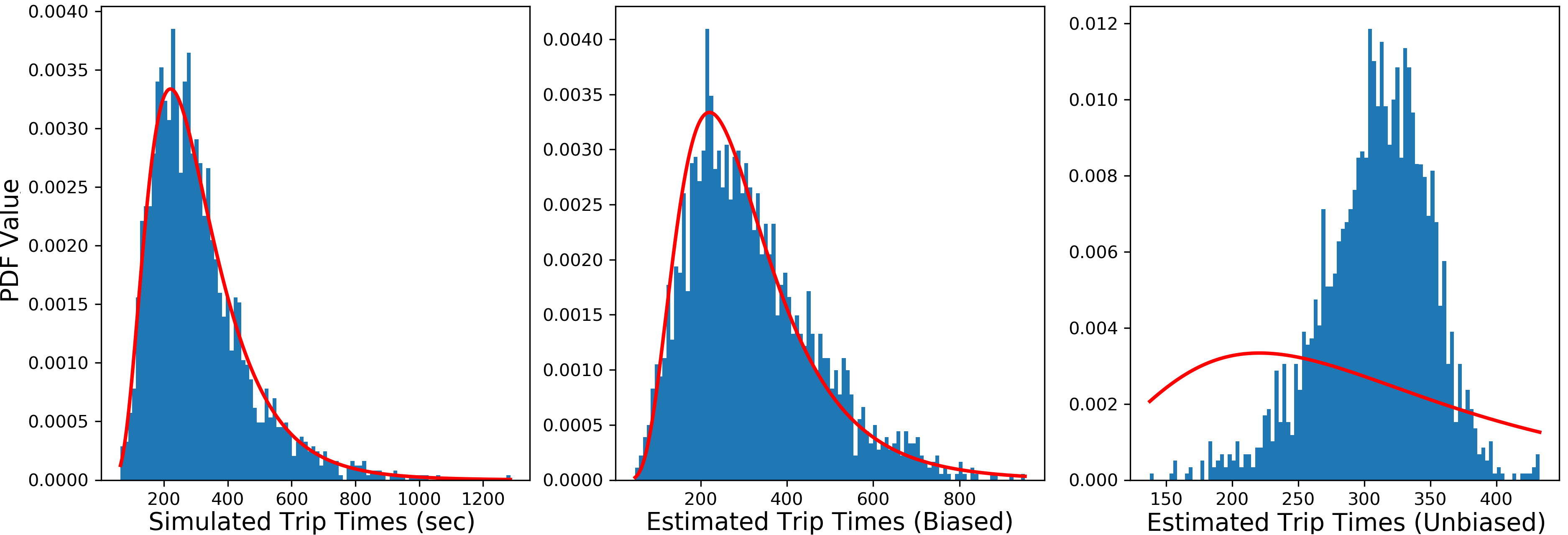}
  \caption{\small Illustrating the effect of biasing travel-time assignments for simulated trips for a TAZ O-D pair with $n_{ij} = 2000$ trips. (Left) Distribution of simulated travel times $y_s$ according to the log-normal distribution drawn in red. (Center) Distribution of estimated trip travel times $S^Tt \approx y$ computed using biased sampling $y(\ell_f(k)) = y_s(\ell_s(k))$. (Right) Distribution of estimated trip travel times $S^Tt \approx y$ computed using unbiased sampling $y = y_s$.}
  \label{fig1:timesamp}
\vspace{-8mm}
\end{figure}

In practice, biased travel-time sampling appears to improve results when there is a relatively large number of trips for a given TAZ O--D pair. For example, Fig.~\ref{fig1:timesamp} illustrates the difference between estimated trip travel-time distributions with and without biased travel-time sampling. For 2000 trip times sampled from the log-normal distribution of a given TAZ O--D pair (left), the distribution of estimated trip travel times deviates less from the target distribution (red line, same across panels) when we bias the travel times assigned to each vertex O--D pair (center) than when we do not (right). 

\section{The optimization process}
\label{sec5}
Our approach solves a series of constrained least-squares problems to fit edge travel times to simulated trips (vertex O--D pairs, travel times, and routes) with travel-time statistics consistent with the Uber dataset. For each iteration $k$, we sample $N$ vertex O--D pairs with sampled travel times vector $y_k \in \mathbb{R}^N_{>0}$ and weighted shortest-paths matrix $S_k \in \{0,1\}^{M \times N}$. Our goal is to then estimate edge travel times vector $t \in \mathbb{R}^M_{>0}$ so that $S_k^Tt \approx y_k$ to satisfy our forward model in Eq.~\ref{equn1}. Thus we have a system of $N$ equations with $M$ unknowns where $M$ is the number of road segments (or edges in the graph). The estimates are then updated using a convex combination of the previous estimates and the solution found by minimizing the mean-squared error with constraints on the unknown coefficients, illustrated by Eqs.~\ref{equn2a}-\ref{equn2c} below.
\begin{align}
    \label{equn2a}
    \hat{t} &= \argmin_{\alpha g_k \leq t \leq \beta h_k} \left\| S^T_k t - y_k \right\|_2^2 \\
    \label{equn2b}
    t_{k+1} &= (1-\lambda_k) t_k + \lambda_k \hat{t} \\
    \lambda_{k+1} &= \theta_k \lambda_k
    \label{equn2c}
\end{align}

In our implementation, we initialize $t_0 = f$ with the free-flow estimates and let $\lambda_0 = 1$, we constrain the elements of the vector $\hat{t}$ to be bounded below by $\alpha g_k = 0.8 f$ and above by $\beta h_k = 1.25 t_k$, and we update our weight $\lambda_k$ using the constant $\theta_k$ = 0.9 for all $k$. Our weighted shortest-paths matrix $S_k$ is recomputed each iteration using our current estimates $t_k$ as weights, and our constrained least-squares sub-problems are solved using the {\tt lsq\_linear} function from the well-known Python scientific computing library SciPy. We run the optimizer until the estimates converge (measured by the magnitude of the average change in the solution vector between iterations, Eq.~\ref{eq:conv} with $\delta = 0.01$), or we reach a maximum number of iterations $k_{\max}$. This iterative scheme is similar to the one proposed in \cite{bertsimas2019travel} but with a choice of optimizer that favors scalability in conjunction with a number of heuristics that improve convergence. The workflow of our proposed travel-time estimation algorithm is given in Algorithm~\ref{algTTE}.
\begin{equation}
    \frac{1}{M} \left\| t_k - t_{k-1} \right\|_2 \leq \delta
    \label{eq:conv}
\end{equation}

\begin{algorithm}
\small
\caption{Travel-time Estimation}
\textbf{Input}: Travel time statistics for TAZ O--D pairs; Road network $G=(V,E)$; Number of trips $N$
\newline \textbf{Output}: Estimated travel times along edges $t$ 
\begin{algorithmic}
\STATE Determine number of trips $n_{ij}$ for each TAZ O--D pair (Section \ref{sec:vertsamp})
\STATE Initialize edges with free-flow travel times $t_0 = f$ and $k=0$
% \STATE $t_0$=Free-flow travel time $\forall e \in E$

\WHILE{ Eq.~\ref{eq:conv} not satisfied}
\STATE Sample $N$ vertex O--D pairs
\STATE Compute shortest-path matrix $S_k$ using weights $t_k$
\STATE Compute $y_k$ using biased travel-time sampling (Algorithm~\ref{alg1}) 
\STATE Compute $t_{k+1}$ according to Eqs.~\ref{equn2a}-\ref{equn2c}
\STATE $k=k+1$
%\STATE Compute errors $\epsilon_k$ using Equation \ref{eq:error}
\ENDWHILE
\end{algorithmic}
\label{algTTE}
\end{algorithm}

% \begin{enumerate}
% \item Split the TAZ O--D pairs into training and test sets. We used a 90-10 split.
% \item Determine the number of simulated trips $n_{ij}$ for each TAZ O--D pair $(i,j) \in \mathcal{U}_k$ based on the TAZ sizes and total number of trips per iteration $N$ (Section \ref{sec:vertsamp}).
% \item For each iteration $k$, sample $n_{ij}$ vertex O--D pairs and biased travel times $y_k$ for each TAZ O--D pair $(i,j) \in \mathcal{U}_k$ in the training and test subsets (Algorithm \ref{alg1}). 
% \item Compute the shortest paths between vertex O--D pairs weighted by the current edge travel-time estimates $t_k$, represented by the shortest-path matrix $S_k$.
% \item Update the estimated edge travel times $t_k$ using a convex combination of the previous iterate and the solution to the constrained least-squares problem (Equations \ref{equn2a}-\ref{equn2c}). 
% \item Calculate prediction errors on the training and test TAZ O--D pairs by comparing the estimated geometric means and those in the Uber dataset (Equation \ref{eq:error}). 
% \item Repeat steps (2) to (6) until convergence (Equation \ref{eq:conv}).
% \end{enumerate}
\section{Experimental results}
\label{sec6}
\begin{figure}[htbp]
  \centering
  \includegraphics[width=0.8\linewidth]{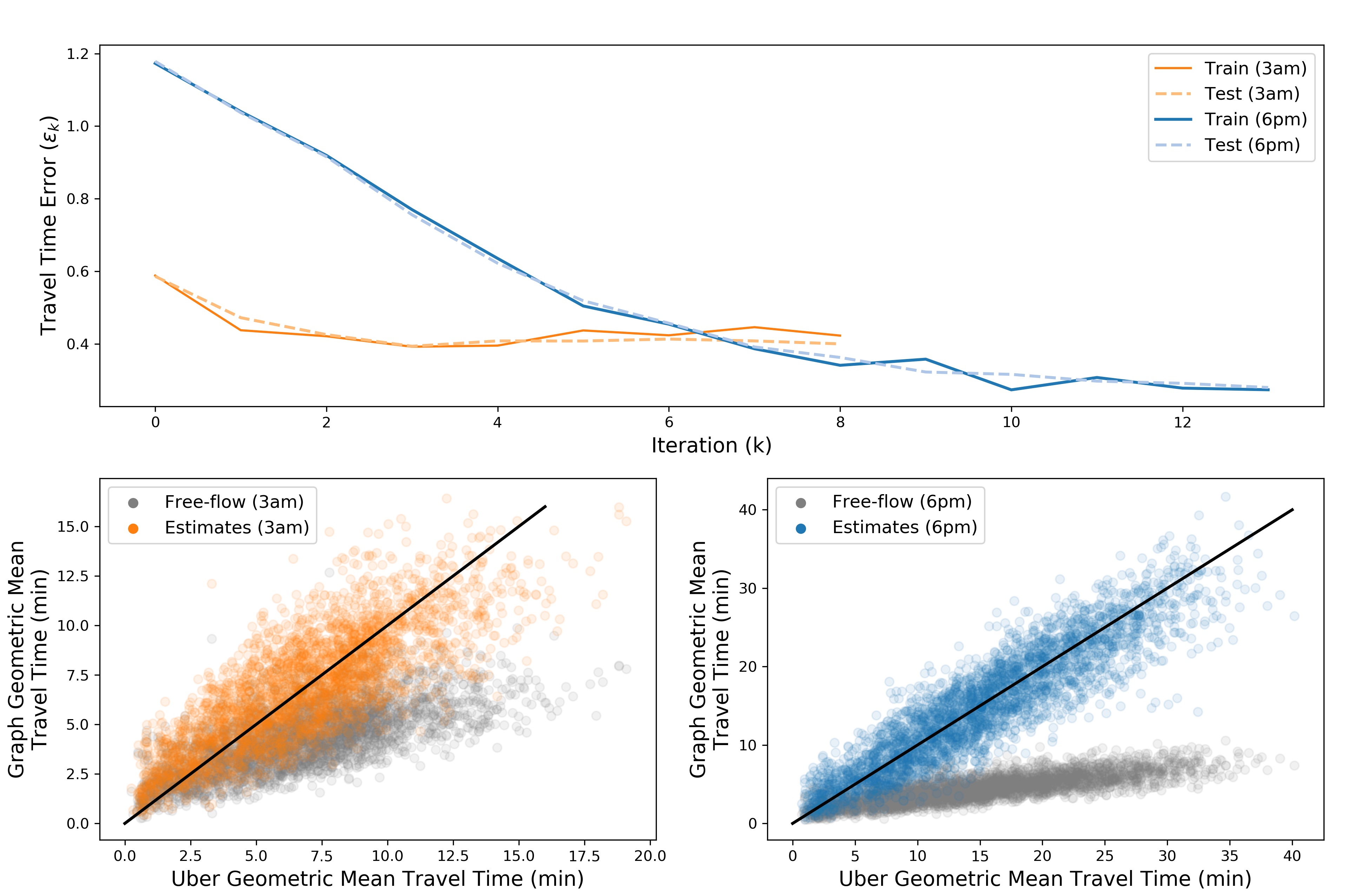}
  \caption{\small Travel-time estimation results for downtown LA for off-peak (3am) and peak traffic (6pm). Top: Convergence of travel time error $\epsilon_k$. Bottom: Geometric mean travel times $g_{ij}$ calculated using free-flow edge weights and estimated edge weights compared to ground-truth data $G_{ij}$ from Uber.}
  \label{fig2:conv}
  
\end{figure}

\begin{figure}[htbp]
  \centering
  \includegraphics[width=0.375\linewidth]{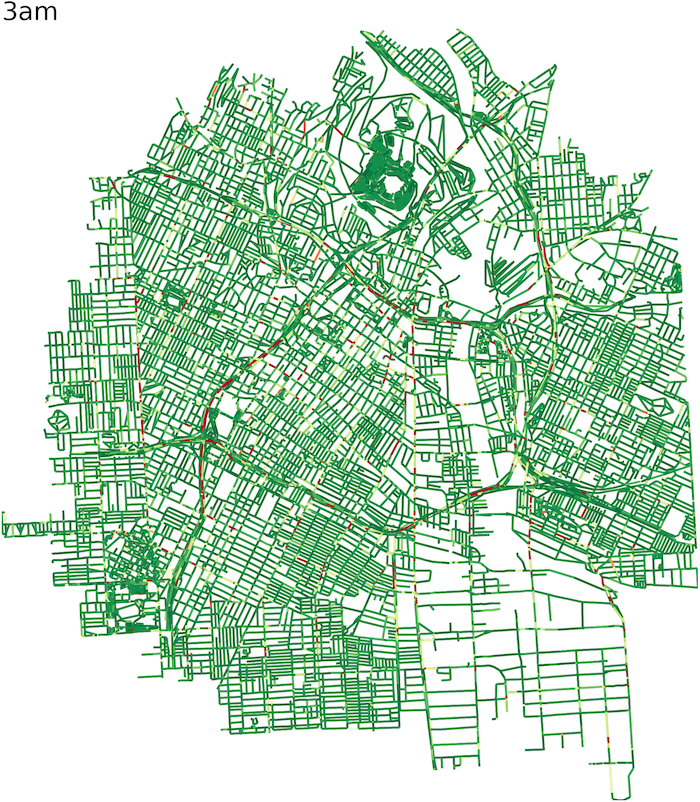}
  \includegraphics[width=0.375\linewidth]{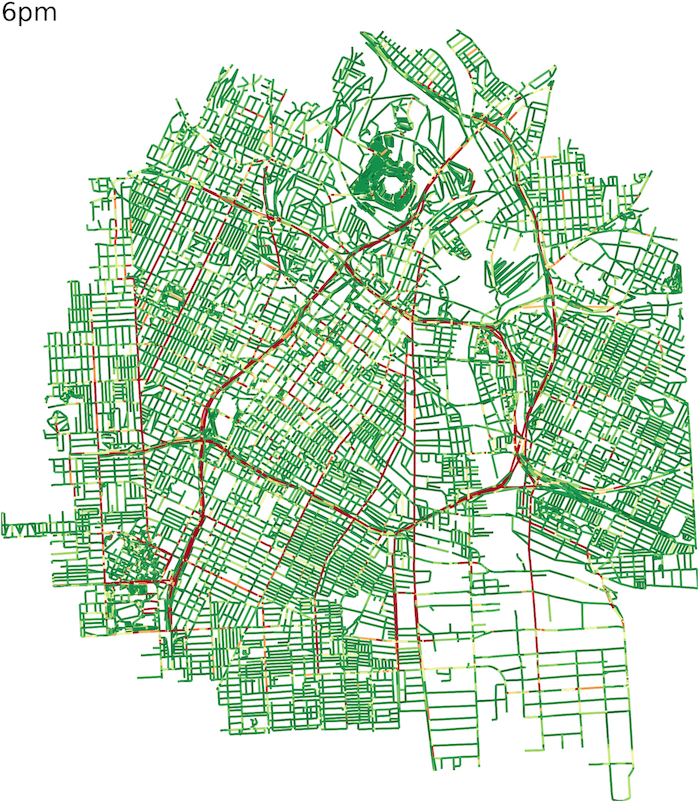}
  \includegraphics[width=0.08\linewidth]{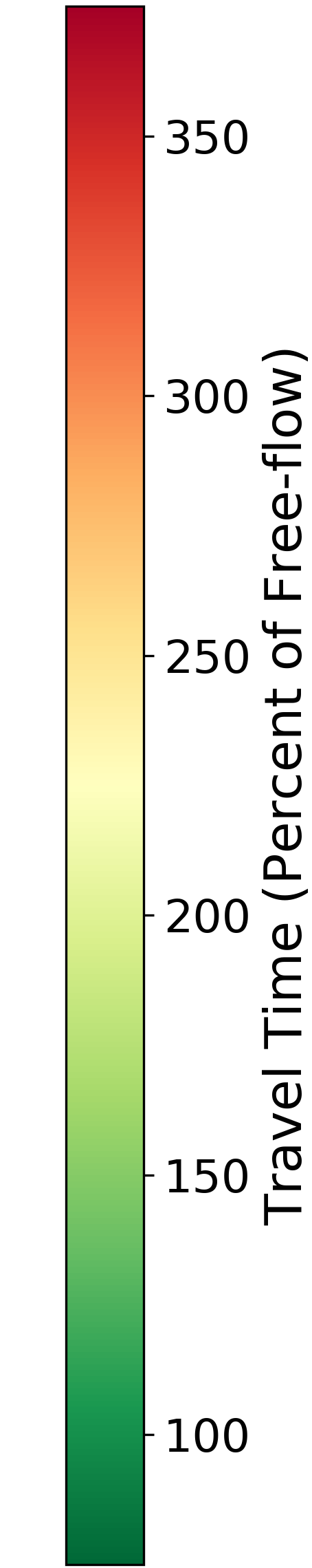}
  \caption{\small Estimated travel time $t$ as a percent of free-flow travel time $f$ in downtown LA for off-peak traffic (3am, left) and peak traffic (6pm, right).}
  \label{fig3:maps}
  \vspace{-5mm}
\end{figure}

We implement the workflow Algorithm~\ref{algTTE} on a road network encompassing a three mile radius of downtown Los Angeles (network \textbf{LA\_DT+3} in Table~\ref{tab1:comp_table}). Fig.~\ref{fig2:conv} shows the convergence of the travel-time error (both training and test sets) for the road network at 3am and 6pm, measured by
\begin{equation}
    \epsilon_k = \sqrt{\frac{1}{N} \sum_{(i,j) \in \mathcal{U}_k} n_{ij}\big( \log g_{ij}(t_k) - \log G_{ij}\big)^2},
    \label{eq:error}
\end{equation}
where $N$ is the total number of trips in the training (test) data, $n_{ij}$ is the number of trips for TAZ O--D pair $(i,j) \in \mathcal{U}_k$ in the training (test) subset, $g_{ij}(t_k)$ is the current estimated geometric mean travel time (Eq.~\ref{eqgmtt}), and $G_{ij}$ is the geometric mean travel-time from the Uber Movement dataset. This RMSLE error metric represents the relative mean-squared error in the logarithm of the estimated geometric mean travel time \cite{bertsimas2019travel}. Fig.~\ref{fig3:maps} shows the estimated travel times (as a percent of free-flow travel time $f$) for both 3am and 6pm, where congested regions (red) are clearly visible in the 6pm plot.

\begin{figure}[htbp]
  \centering
  \includegraphics[width=0.8\linewidth]{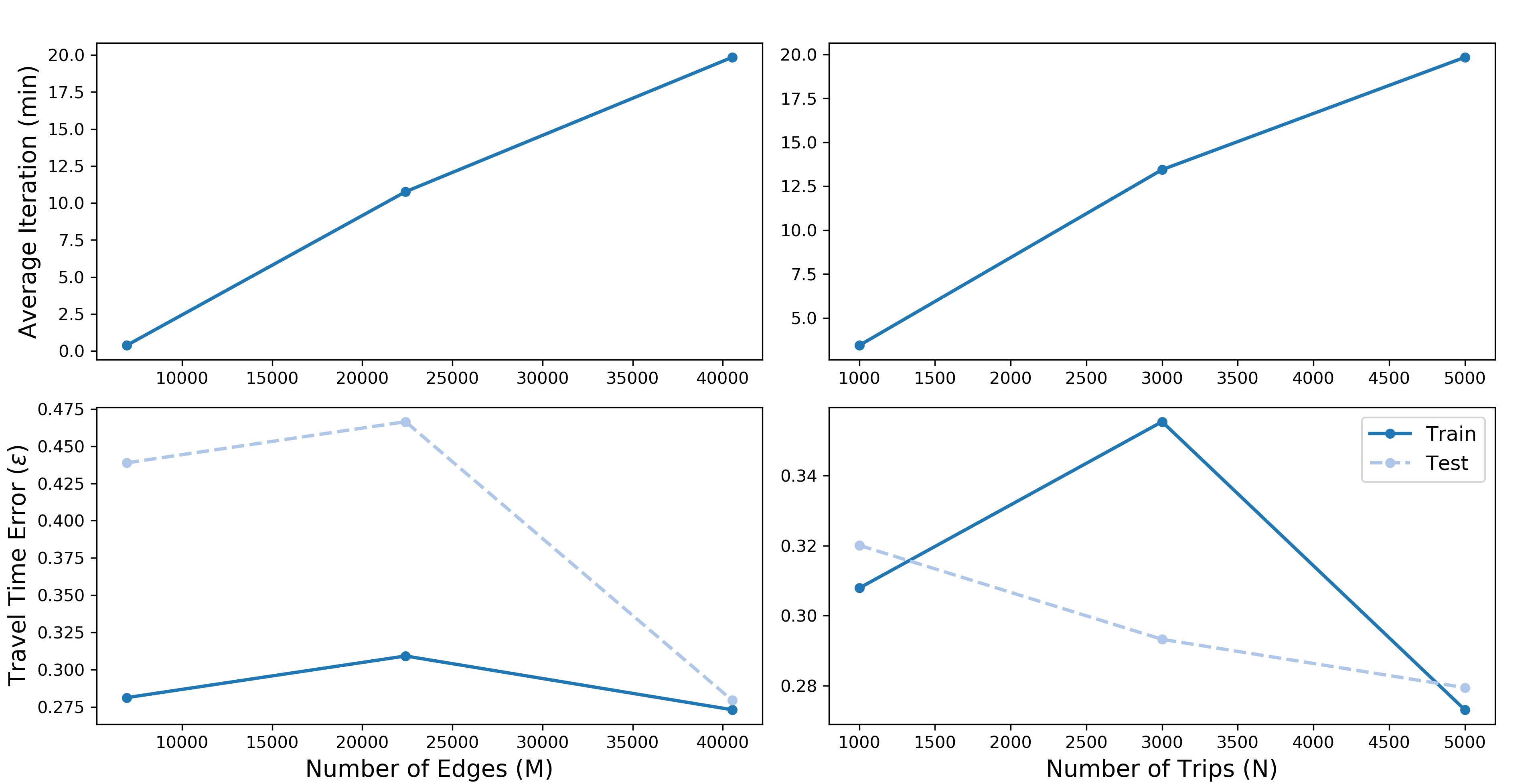}
  \caption{\small Scaling results for travel time estimation in downtown LA for peak traffic (6pm). Left column: Results for one, two, and three mile radii of downtown. Right column: Results for 1000, 3000, and 5000 simulated trips per iteration. Top row: The average time per iteration scales proportionally with both the number of edges and number of trips. Bottom row: The travel time errors for both training and test sets achieve their lowest values for larger problem sizes.}
  \label{fig4:scaling1}
  \vspace{-8mm}
\end{figure}

\section{Scalability and graph sparsification}
\label{sec7}
The computational complexity of our proposed model is proportional to the number of trips $N$ and the number of the edges $M$ (Fig. \ref{fig4:scaling1}). Due to the amount of data available, including $368,419$ vertices and $905,622$ edges in the full LA road network graph and between $507,449$ (3am) and $1,115,432$ (6pm) TAZ O--D pairs in the 2019 Quarter 1 Weekdays-Only Uber dataset, the constrained least-squares optimizer becomes a computational bottleneck, limiting our ability to solve large systems (i.e., street-level travel-time estimation of a large geographical area). Since our goal is to generate a data-driven solution, we aim to use as much data as possible, therefore leading us to systemically reduce the number of unknowns. One strategy would be to sparsify the graph by dropping edges according to a pre-determined criterion but still maintaining the graph connectivity. This strategy, however, changes routing, which can result in skewed travel-time estimations. For example, consider two paths between vertex $u$ and $v$: the shortest route through intermediate vertices $s$ and $t$, and an alternative, but longer, route through vertices $x$ and $y$. If a sparsification algorithm removes the road segment between $s$ and $t$ based on some metric (e.g., low edge betweenness centrality), our model would incorrectly estimate the travel time from $u$ to $v$ using the available road segment through $x$ and $y$. 

In light of this, our approach is to pseudo-sparsify the underlying road network. Rather than remove edges from the graph, we sort road segments into two sets based on their significance with respect to traffic flow (e.g., betweenness centrality). We then estimate the travel-time along the edges with the top $p\%$ significance, setting the travel times along the remaining edges to the free-flow travel times. In our implementation, we sort edges based on their betweenness centrality, computed using shortest-paths weighted by free-flow travel times. For instance, if $i$ are the indices of the edges with the top $p\%$ betweenness and $j$ are indices of the remaining edges, we solve a modified version of Eqs. \ref{equn2a}-\ref{equn2b} each iteration:
\begin{align}
    \hat{t} &= \argmin_{\alpha g_k(i) \leq t \leq \beta h_k(i)} \left\| S_k(i)^T t + S_k(j)^T f(j) - y_k \right\|_2^2 \\
    t_{k+1}(i) &= (1-\lambda_k) t_k(i) + \lambda_k \hat{t}, \quad t_{k+1}(j) = f(j)
\end{align}
In this way, we can control the number of unknowns $M$ and scale our method for larger geographical areas.

% \begin{algorithm}
% \small
% \caption{Graph pseudo-sparsification}
% %\newline \textbf{Input}: Graph $G$; percentage of edges for travel time estimation $p$.
% %\newline \textbf{Output}: Graph $G$ where $(1-p)\%$ edges with pre-computed estimation }
% \begin{algorithmic}
% \STATE{$m$=number\_of\_ edges($G$)}
% \STATE{$k$ = $m*(1-p)$}
% \STATE{Compute edge betweenness centrality of $G$}
% \STATE{$ES$=$k$-smallest links (edges) w.r.t betweenness}
% \FOR{$e$ in $ES$}
%     \STATE $l$=length($e$)
%     \STATE $s$=posted\_speed\_limit($e$)
%     \STATE free flow travel time, $t$=$l/s$
%     \STATE{$e[travel\_time]=t$}
% \ENDFOR
% \end{algorithmic}
% \end{algorithm}

\begin{figure}[htbp]
  \centering
  \includegraphics[width=0.8\linewidth]{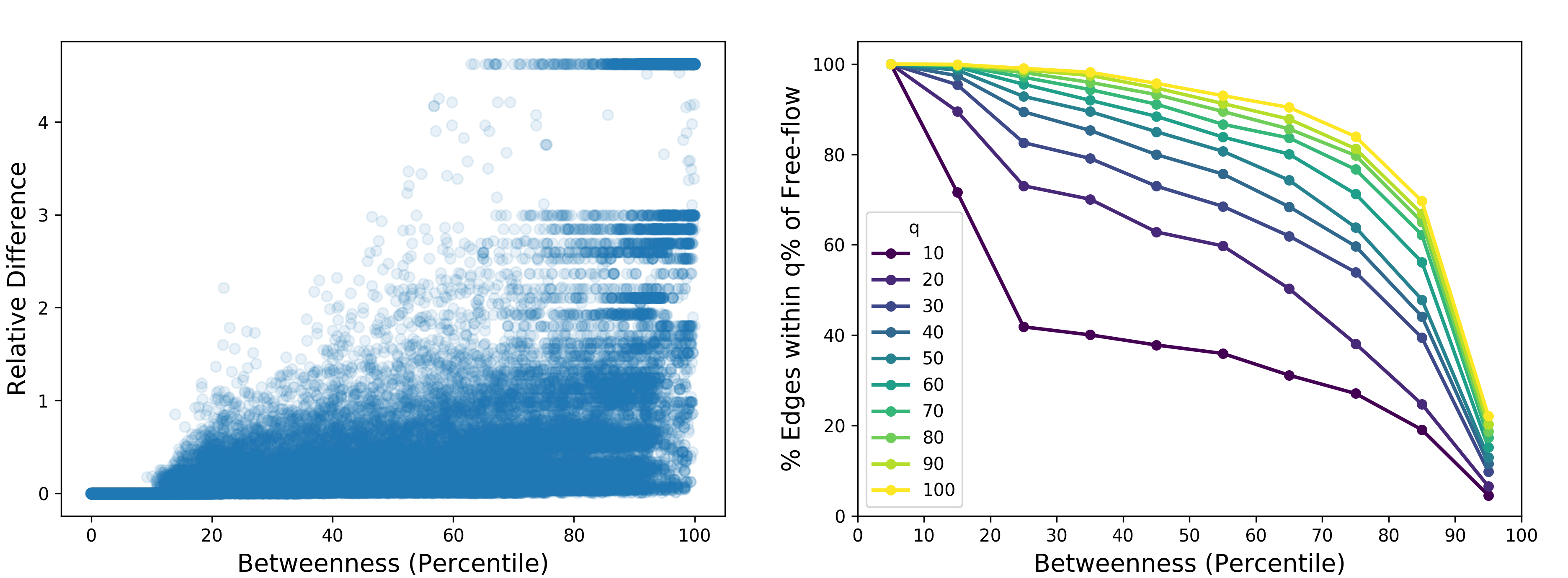}
  \caption{\small Relative difference ($|t - f|/f$) between estimated travel time $t$ and free-flow travel time $f$ by percentile of betweenness for peak downtown LA traffic (left) and fraction of estimated edge travel times that are within $q\%$ of free-flow ($|t-f|/f\leq q\%$), where edges are binned by percentile (right). Edge betweenness centrality is calculated using shortest paths weighted by free-flow travel time. }
  \label{fig4:correlation}
  \vspace{-5mm}
\end{figure}

\begin{figure}[htbp]
  \centering
  \includegraphics[width=0.8\linewidth]{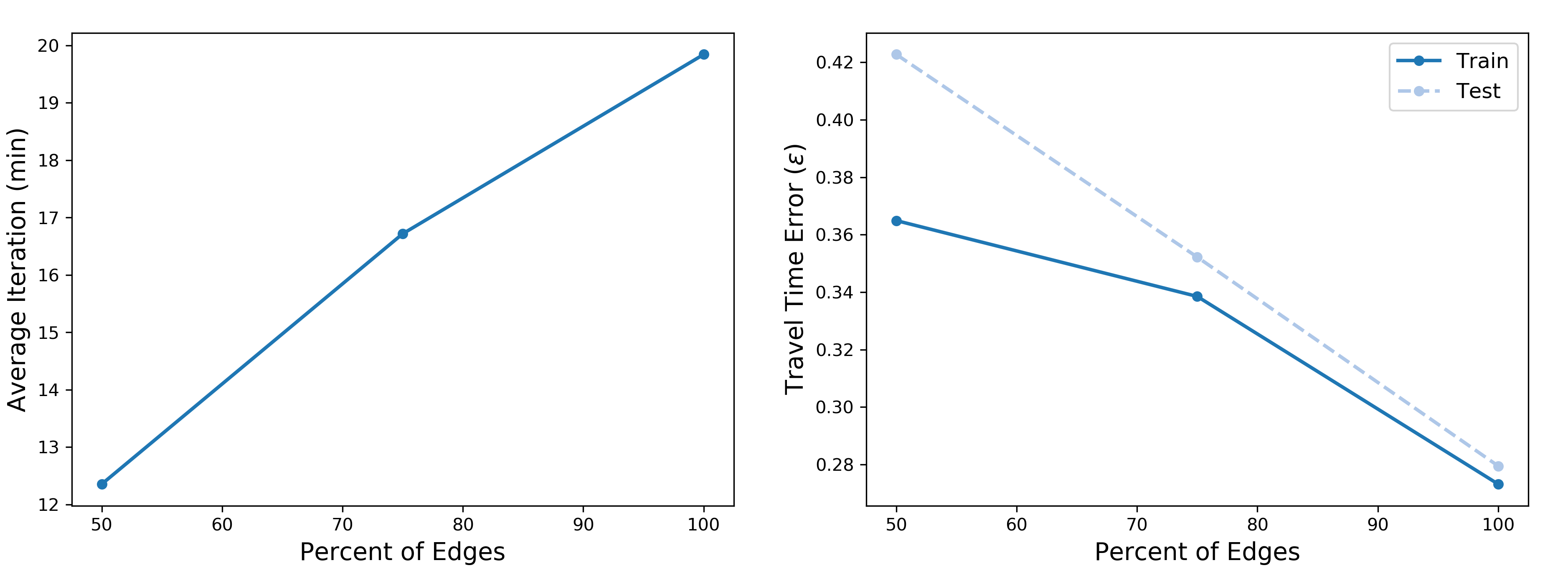}
  \caption{\small Scaling results for travel time estimation in downtown LA for peak traffic (6pm) solved using pseudo-sparsified graphs, where 50\%, 75\%, and 100\% of the edges were estimated and the remainder set to free-flow travel times. Left: The average time per iteration scales proportionally with the percent of edges estimated. Right: Travel time errors decrease as less edges are set to free-flow travel time.}
  \label{fig5:scaling2}
  \vspace{-5mm}
\end{figure}

We test our pseudo-sparsification approach using different values of $p$ (Figs.~\ref{fig4:correlation}-\ref{fig5:scaling2}). In Fig.~\ref{fig4:correlation} (left), we first run our algorithm on the full graph (i.e., $p = 100\%$), and plot the relative difference between our estimated travel times and the initial free-flow travel times $|t - f|/f$ versus the percentile of the edges sorted by betweenness centrality. We observe that roughly $15\%$ of lowest betweenness edges retain their free-flow travel time as the final estimate, suggesting that we can set these edges to their free-flow travel times to reduce the number of unknowns.  

Of course, drivers rarely maintain the exact speed limit. To account for this uncertainty in free-flow travel times, we sort the edges into bins based on their betweenness percentiles and plot the fraction of edges within each bin that have a relative difference within a factor of $q\%$ of their free-flow travel times (Fig.~\ref{fig4:correlation}, right). We observe that the high betweenness edges are less likely to have an estimated travel time close to their free-flow times irrespective of the level of uncertainty $q$. Therefore, we can pseudo-sparsify the graph to different extents depending upon the level of uncertainty that can be tolerated.

In Fig.~\ref{fig5:scaling2}, we vary the number of edges by estimating only the top 50\%, 75\%, and 100\% of edges by betweenness. Here the time per iteration scales proportionally with the number of unknown edges included in the problem, and the travel-time error given by Eq.~\ref{eq:error} increases as we assign more edges to their free-flow travel times. 

%%%%%%%%%%%%%%%%%%%%

\section{Related work}
\label{sec2}
Urban traffic modeling has seen a recent surge of interest in the use of graph analytics and machine learning methods. Reference \cite{vlahogianni2014short} provides an overview of many of these methods. In this section we summarize the prior work related to both of these themes as applied to road transportation networks.

The authors in \cite{tian2016analysis} use graph analytics on networks with three different weighting schemes to perform a statistical characterization of the Beijing road network. Many prior publications consider betweenness centrality \cite{freeman1977set} to be an important metric when applied to road networks, as it is argued to be a direct predictor of important links in urban transport. It has been shown that betweenness centrality is highly correlated with the traffic flow count on a road network \cite{crucitti2006centrality,jiang2009street,porta2006network}, a natural result of including travel time as a factor when selecting trip routes. In real-world scenarios, however, route choices are also influenced by time-of-day and other socio-economic factors. Using these observations, the authors in \cite{puzis2013augmented} define an augmented betweenness centrality measure where shortest paths are weighted according to a traffic demand model based on census tracts and traffic analysis zones. The authors show that this new centrality measure correlates better with traffic flow than other centrality measures. Similarly, in \cite{cheng2015measuring} the authors employ analytics in the form of novel graph centrality measures to derive insights into the traffic flow patters in Singapore. Graph models, along with heterogeneous data sources, were leveraged to understand the urban traffic patterns in \cite{oberoi2017spatial}. Finally, the authors in \cite{gundaliya2008heterogeneous} utilize a grid-based fabric and cellular automata for modeling arterial traffic, resulting in gains in computational efficiency. 

Many approaches make use of machine learning models and optimization methods to model various aspects of urban traffic flow. In \cite{li2018diffusion}, the authors leverage a deep-learning approach in the form of a diffusion convolutional recurrent neural network (DCRNN) to forecast short-term freeway traffic counts in the LA and San Francisco Bay Area networks. The authors in \cite{cheng2018deeptransport} also propose a deep-learning approach that brings together convolutional neural networks and recurrent neural networks with long short-term memory (LSTM) units, utilizing their architecture for short-term traffic count extrapolation at 349 locations on the Beijing road network. In a set of articles \cite{rakha2015use,zhang2015arterial}, the authors leverage data from Bluetooth and GPS probe sensors for travel-time estimation and validation. Coupled hidden Markov models (CHMM) were used in \cite{herring2010estimating} to model the evolution of traffic states, applied to a sparse taxi-fleet dataset for the San Francisco Bay area road network. In a subsequent publication \cite{hofleitner2012learning} leveraging the same dataset, the authors employ a dynamic Bayesian network to learn arterial dynamics. 

\section{Conclusions and future work}
\label{sec8}
In this work we leveraged coarse-grained Uber Movement data in the form of TAZ O--D pair summary statistics to provide estimates of fine-grained, street-level travel times. Our techniques for trip simulation and biased travel-time sampling were used in conjunction with weighted shortest-path routing to set up a system of linear equations with unknown edge travel times. The travel times were iteratively refined using a constrained least-squares optimizer for multiple batches of simulated trips. In our largest road network (40,522 edges and 24,472 TAZ O--D pairs), we achieved a RMSLE of 0.28 on our test set after 13 iterations with 4.7 hours total runtime (32-core Intel Xeon E7-8860 system with 2.27 GHz processor speed and 512 GB primary memory). We demonstrated a graph pseudo-sparsification algorithm in order to improve the computational efficiency of our estimation routines. Our future work involves improving the optimization process, formalizing the sparsification approach, and scaling the approach to metropolitan-sized road networks using high-performance computing systems. 

%\section*{Acknowledgments}
%This report and the work described were sponsored by the U.S. Department of Energy (DOE) Vehicle Technologies Office (VTO) under the Big
%Data Solutions for Mobility Program, an initiative of the Energy Efficient
%Mobility Systems (EEMS) Program. The following DOE Office of Energy
%Efficiency and Renewable Energy (EERE) managers played important roles
%in establishing the project concept, advancing implementation, and providing
%ongoing guidance: David Anderson and Prasad Gupte.

% We implement the proposed algorithms in Python $(version 3.7)$.

% All the experiments are conducted on an $32$-core Intel Xeon $E7-8860$ system with processor speed of $2.27$ GHz. The system has $512$ Giga Byte primary memory.

% Test set for 6pm
% Initial (free-flow) RMSLE: 1.1786117525467117
% Final RMSLE: 0.27950243463870017
% Difference: 0.89910931790801152
% Percent decrease: 76.285453285633793

% We showed that these result in the RMSLE of the geometric mean travel time over the TAZ O--D pairs in the test data being reduced by 76.3\% in our largest road network (40,522 edges and 24,472 TAZ O--D pairs) with a 4.7 hours total runtime.

\bibliographystyle{siamplain}

\end{document}